\documentclass[runningheads]{llncs}

\usepackage{amsmath}
\DeclareMathOperator*{\argmin}{arg\,min}
\usepackage{amssymb}
\usepackage{graphicx}
\usepackage{booktabs}
\usepackage{multirow}

\begin{document}

\title{Root-Cause Analysis of Activation Cascade Differences in Brain Networks}

\titlerunning{Root-Cause Analysis of Activation Cascade Differences in Brain Networks}

\author{Qihang Yao\inst{}\orcidID{0000-0001-5091-3850} \and
Manoj Chandrasekaran\inst{}\orcidID{0000-0002-8625-0013} \and
Constantine Dovrolis\inst{}\orcidID{0000-0002-4491-861X}}

\institute{Georgia Institute of Technology, Atlanta GA 30332, USA
\email{\{qihang.yao,manojc,constantine\}@gatech.edu}}
\authorrunning{Q. Yao et al.}

\maketitle

\begin{abstract}
Diffusion MRI imaging and tractography algorithms have enabled the mapping of the macro-scale connectome of the entire brain.
At the functional level, probably the simplest way to study the dynamics of macro-scale brain activity is to compute the ``activation cascade'' that follows the artificial stimulation of a source region. 
Such cascades can be computed using the Linear Threshold model on a weighted graph representation of the connectome. 
The question we focus on is: {\em if we are given such activation cascades for two groups, say A and B (e.g. Controls versus a mental disorder), what is the smallest set of brain connectivity (graph edge weight) changes that are sufficient to explain the observed differences in the activation cascades between the two groups?} 
We have developed and computationally validated an efficient algorithm, TRACED, to solve the previous problem. 
We argue that this approach to compare the connectomes of two groups, based on activation cascades, is more insightful than simply identifying "static" network differences (such as edges with large weight or centrality differences).  
We have also applied the proposed method in the comparison between a Major Depressive Disorder (MDD) group versus healthy controls and briefly report the resulting set of connections that cause most of the observed cascade differences.

\end{abstract}

% \newpage
\section{Introduction}
\label{sec:intro}

Diffusion MRI imaging and tractography algorithms have enabled the mapping of the macro-scale connectome of the entire brain~\cite{sporns2005connectome}. This network representation enables the application of powerful tools from graph theory and graph algorithms in the study of the brain's structure and function. Earlier work has focused on various important network properties of the brain such as small worldness~\cite{bassett2017small}, presence of hubs~\cite{hwang2013development}, modularity~\cite{sporns2016modular}, etc. These studies have revealed that seemingly local pathologies in specific regions can have far-reaching global effects on other parts of the brain~\cite{stam2014modern,rehme2013cerebral}. 

Probably the simplest way to study the dynamics of brain activity at the macro-scale is to compute the ``activation cascade'' that is generated by the artificial stimulation of a source region. Activation cascades, represented in the form of directed acyclic graphs (DAGs), describe how an activation starting from one region (i.e., source node) propagates to the rest of the brain, activating other brain regions along the way. Previous work has applied the Asynchronous Linear Threshold (ALT) model on the mouse meso-scale connectome to simulate the propagation and integration of sensory signals through activation cascades~\cite{shadi2020multisensory}. Those modeling results were validated with functional data from cortical voltage-sensitive dye imaging, showing that the order of node activations in the model matches quite well with the empirical activation order observed experimentally~\cite{shadi2020multisensory}.

The question that we focus on in this study is: suppose we are given two groups with significant differences in the activation cascades generated in their brain networks. What is the smallest set of brain connectivity (i.e., graph edge weight) changes that are sufficient to explain the observed differences in the activation cascades between the two groups? Answering this question can be valuable in many studies when two groups should be compared, not only in terms of structural connectome differences, but also in terms of functional dynamics. For example, we can identify a (generally small) set of brain connectivity changes that appear to cause the functional activation differences in a given disorder, by comparing the corresponding activation cascades with healthy controls. Further, the corresponding connections can be used as possible targets in interventions and treatments such as deep brain stimulation~\cite{van2015evidence,riva2018connectomic}.

We have developed an algorithm named TRACED (The Root-cause of Activation Cascade Differences) to solve the previous problem. TRACED starts by identifying node membership differences between the two groups (say A and B) within the activation cascade of each source. Then, for each source, we identify the smallest set of edges that, if their weights in group A are modified to be equal to the weights in group B, the corresponding activation cascades will be the same in both groups. We have computationally validated TRACED across many test cases.
Additionally, we have applied TRACED in the comparison between a group of patients with major depressive disorder (MDD) and a group of controls. This paper focuses on the proposed computational method -- a more comprehensive MDD-focused study of the two groups will be presented in a different article. 

Previous work detected significant topological differences in terms of network metrics such as edge weights and centrality measures for various neurological disorders, including multiple sclerosis~\cite{fleischer2019graph,llufriu2017structural}, Alzheimer's disease~\cite{fischer2015altered}, Parkinson's disease~\cite{wen2017structural}, and schizophrenia~\cite{Fornito2012schizophrenia}. We argue that the activation cascade approach to comparing the connectomes of two groups is more insightful than simply identifying such "static" network differences. The former makes some clear and simple assumptions about the processing and propagation of information in the brain, and it creates a causal connection between structural changes and functional effects. Therefore, the identified abnormalities are more interpretable and robust to subject variability.

\section{Linear threshold model and activation cascades}
\label{sec:lt-model-cascade}

Our starting point is a structural macro-scale brain network. In this network representation, the graph is denoted by $G = (V, E)$, each node in $V$ corresponds to a brain region, and $E$ contains edges that correspond to connectivity between brain regions. For DTI-based structural networks, the edges are undirected. Each edge $(x, y)$ in $E$ is associated with a weight $w(x, y)$ that represents the strength of the corresponding connection.

% \subsection{Linear threshold model}
% \label{sec:lt-model}

In the linear threshold model, each node can be either active or inactive. Initially, all nodes are inactive, except a single source node. If a neighbor $y$ of a node $x$ is active, then we say that $x$ ``receives an activation'' from $y$ with strength $w(y, x)$. Node $x$ becomes active if it receives a cumulative activation from all its active neighbors that is more than a threshold $\theta$. 

More formally, a node $x$ at time $t$ is associated with a binary state variable $A(x, t)$ indicating whether $x$ is active (1) or not (0). 
For the source node $s$, we have that $A(s, t=0) = 1$ and for all other nodes:
\begin{equation}
  A(x, t+1) = 
  1 \text{ if } \sum_{y \mid (y, x) \in E} {w(y, x) A(y, t)} \geq \theta \\
\end{equation}
for $t\geq 0$. If $x$ becomes active in the cascade of source $s$, $t_s(x)$ is the time of its activation. 
By convention, $t_s(x) = \infty$ if node $x$ never gets active. 

% \subsection{Cascade and membership difference}
% \label{sec:membership-difference}

\begin{figure}
    \centering
    \includegraphics[width=\textwidth]{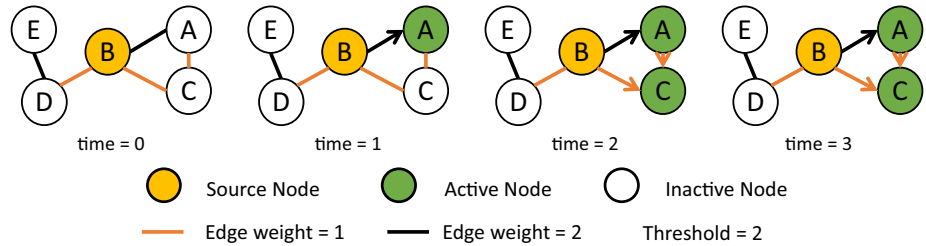}
    \caption{An illustrative example of an activation cascade obtained using the linear threshold model. Node B is the source of the cascade. The threshold $\theta=2$. Node A gets active through the edge (A, B), and node C becomes active after both A and B are active. The rest of the nodes stay inactive in this cascade. }
    \label{fig:activation-cascade}
\end{figure}

An activation cascade, in the form of a directed acyclic graph (DAG), shows whether as well as how each node becomes active. The nodes in the activation cascade of source $s$ form the following set:
\begin{equation}
  U(s) = \left\{ x\in V \mid  t_s(x) < \infty \right\}
\end{equation} 
The edges in the activation cascade include $(x, y) \in E$ if node $x$ becomes active before $y$. So, the presence of this edge in the cascade DAG means that $x$ participates in the activation of $y$. Mathematically, 
\begin{equation}
  F(s) = \{(x, y) \in E \mid  t_s(y) < t_s(x) \}
\end{equation}
We denote the activation cascade as $H(s)=\{U(s), F(s)\}$. In Fig.~\ref{fig:activation-cascade} we show a simple example illustrating an activation cascade generated in a toy network using the linear threshold model.

For a given $\theta$, different source nodes may give different cascade sizes. Some source nodes do not activate any other node giving rise to \textit{empty cascades}, while other source nodes may activate every node in the network causing a \textit{full cascade}. The third case is that of a \textit{partial cascade}, which is more likely in practice. It would be unrealistic to set the threshold $\theta$ so high that we get many empty cascades -- that would correspond to a comatose brain! However, it would also be unrealistic to set $\theta$ so low that we get many full cascades. The previous observations guide us to choose a range of $\theta$ values that result in more partial cascades, across different source nodes.

When comparing the structural brain networks of two subjects, or two groups, we rely on the membership of each source's cascade: 
If a node $x$ is active in the cascade of source $s$ in one network, is $x$ also active in the corresponding cascade of the other network?
The similarity between the node membership of two cascades is quantified using the Jaccard similarity metric, 
applied on the set of active nodes in the two cascades. A small Jaccard similarity represents a large difference between the two cascades. Therefore, using $U(s)$ and $U'(s)$ to denote the set of active nodes in networks $G$ and $G'$, respectively, after the activation of source $s$, the difference between the two cascades is quantified by:
\begin{equation}
    d\{U(s), U'(s)\} = 1 - J\{U(s), U'(s)\}  = 1 - \frac{|U(s) \cap U'(s)|}{|U(s) \cup U'(s)|}
\end{equation}
where $J\{U(s), U'(s)\}$ is the Jaccard similarity of the two cascades.

\section{TRACED algorithm}
\label{sec:root-cause}

\begin{figure}[ht!]
    \centering
    \includegraphics[width=0.9\textwidth]{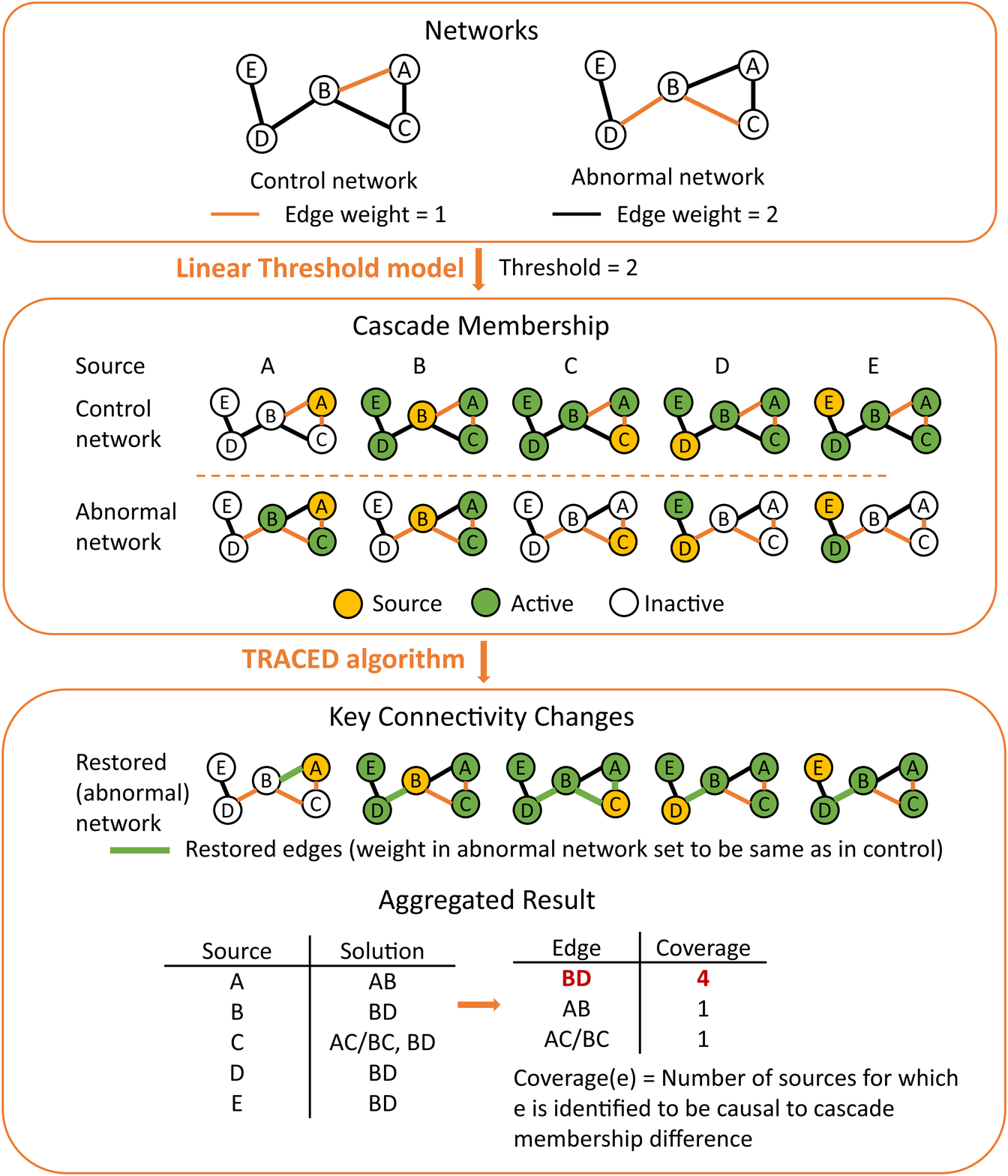}
    \caption{Method overview: The abnormal and control networks may have several edges with different weights. We generate the activation cascade for each source using the linear threshold model, and identify the cascade membership differences across the two networks. Then we identify a subset of edges (containing only edge BD in this example) whose weight change can explain the majority of the observed cascade differences. In other words, if we restore the weights of this subset of edges in the abnormal network to be equal to the corresponding weights in the control network, the majority of the cascade differences between two networks no longer exist. }
    \label{fig:comparison-overview}
\end{figure}

We expect that a mental disorder (or any other genuine distinction in the structural brain networks of a group) would cause cascade membership differences for several different sources~\cite{zeng2012identifying,stam2007small}. Additionally, it is reasonable to expect that these cascade membership differences will be caused by a rather small set of brain connectivity abnormalities (a larger set of abnormalities would probably be lethal). Under these assumptions, we aim to detect the smallest set of edge weight changes that can explain the observed cascade membership differences between the two groups.

\paragraph{{\bf The case of a single source node:}}
% \subsubsection{Problem Formulation}
The problem of finding the root-cause for the activation cascade differences of a single source $s$ can be formulated as follows: We are given the cascade of $s$ in the control and the abnormal networks. {\em Compute the minimum set of edges $C$ in the abnormal network so that, if we restore the weights of those edges to be equal to the corresponding weights in the control network, the activation cascade of $s$ will be identical in the two networks.} The $C$-restored network is created by replacing the weight of edge $e$ ($e \in C$), in the abnormal network with the weight of $e$ in the control network. 

The mathematical formulation of the previous problem is:
\begin{equation}
    \hat{C} = \argmin_{C \in \{E \cup E'\}} |C| \ \text{s.t.}\ U'_C(s) = U(s)
\end{equation}
where the set of active nodes in the control cascade of $s$ is denoted by $U(s)$, the set of active nodes in the abnormal cascade of $s$ is  $U'(s)$, and the set of active nodes in the $C$-restored network of  $s$ is $U'_C(s)$. By convention, we take the weight of any edges that are not present as 0.

% \subsubsection{Algorithm}
A naive algorithm would be to search among all $2^m$ solutions ($m = |E \cup E'|$) but that would be computationally infeasible for the scale of structural brain networks. 

Instead, the TRACED algorithm starts from an empty set $C$ and gradually ``grows'' the solution by adding one edge at a time. The original empty set $C$ can grow into $m$ different sets, each with a distinct edge. In the next step, each of these $m$ sets can include one of the remaining $m - 1$ edges, creating a total of $m(m-1)$ sets with two edges each. This way, when $\hat{C}$ is found, the number of candidate solutions is $m^{k}$, where $k = |\hat{C}|$. Since we are adding edges step by step following an approach similar to breadth-first-search, the solution is guaranteed to be optimal.
Note that even though the run-time of this approach grows exponentially with the solution size $k$, we expect (as previously mentioned) that $k$ will be small in practice. 

The run-time of the algorithm can be improved however based on the following observation. Let us define as ``candidate edges'' the edges that point from $U(s) \cap U'_C(s)$ (nodes active in both cascades) to $U(s) \triangle U'_C(s)$ (nodes active in one cascade but not the other). We know that at each ``growth'' step at least one of the candidate edges should be added to the solution. Otherwise, it is impossible to change the activation status of the nodes in $U(s) \triangle U'_C(s)$. Therefore, in each step we only consider candidate edges, and thus limit the number of new possible solutions created. If $b$ is an upper bound on the number of candidate edges, the number of total solutions generated during the search is at most $b^{k}$.

\begin{figure}[ht!]
    \centering
    \includegraphics[width=\textwidth]{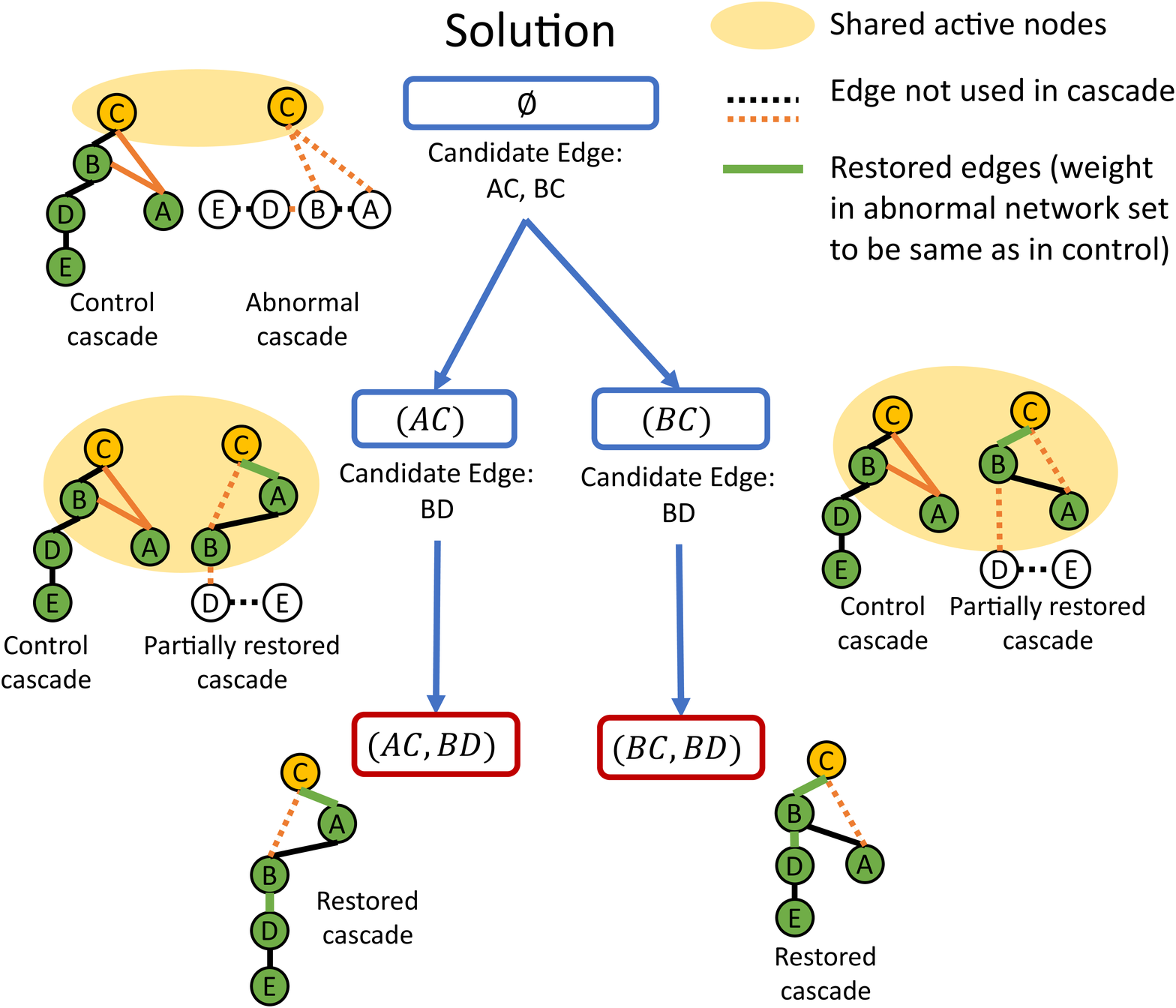}
    \caption{Illustration of TRACED: The tree structure shows how the solution is gradually computed one edge at a time  -- different branches of the tree can lead to different solutions. The final solutions are marked in red. Along with each candidate solution $C$, we present the corresponding cascade $H'_C(s)$. In this example, two solutions can explain equally well the observed differences between the two cascades that originate from source $C$. }
    \label{fig:algorithm-visualization}
\end{figure}

Fig.~\ref{fig:algorithm-visualization} illustrates the execution of the TRACED algorithm with a small example. We start with an empty solution $C$ and with the two activation cascades (control and abnormal) for a single source $s$. Then, we identify the candidate edges between the two cascades. For each candidate edge we ``grow'' a new branch of the solution tree. We repeat these steps until $U(s) = U'_C(s)$. 
% The pseudo-code of this algorithm is as in Algorithm~\ref{alg:iterative-approach}.
TRACED has a time complexity of $O(b^{k} (|V|+|E'|))$ because it iterates through  $b^{k}$ candidate solutions and executes the linear threshold model once for each possible solution. 

In Section~\ref{sec:iterative-Optimization} we introduce an improvement that further reduces the average run-time and allows multiple optimal solutions to be found, by adding more than one edge into a candidate solution at each step. That improvement does not change the algorithm's main idea or its worst-case run time.

To computationally validate the correctness of the algorithm, we created pairs of small-scale graphs for which we know the edges that cause activation cascade differences between the two networks. These examples are designed so that they vary in several factors: they can have one or multiple optimal solutions, only one edge or multiple edges in one solution, and edges in a solution that are dependent on each other (i.e. an edge included in the cascades only when the weight of another edge is restored). TRACED results in the correct results in all cases, identifying one or multiple optimal solutions correctly.

\paragraph{{\bf Aggregation across different source nodes:}} 
The previous algorithm may produce different sets of edges for different source nodes. Some of these edges may be the result of noise in the data or other artifacts. We select a subset of these edges based on the following argument: if TRACED identifies a certain edge as causal, not only for one source but for multiple, it is likely that that edge represents a genuine and important difference between the control and abnormal networks. 

We use the \textit{coverage} metric to measure the number of sources for which an edge $e$ has been identified as causal for the cascade membership differences. Edges with higher \textit{coverage} play a more central role in the observed differences between the two networks.

% how we use the coverage metric. 
% how the OVERALL algorithm eventually chooses which edges to report.
To test if the \textit{coverage} of an edge is significant or not, we construct a null hypothesis that all edges in the network have the same probability ($\frac{|\hat{C}(s)|}{|E|}$, where $\hat{C}(s)$ refers to the set of edges identified to be causal to cascade membership differences with source node $s$) to be reported as causal for source $s$. 
Under that assumption, the \textit{coverage} metric follows a  binomial distribution:
\begin{equation}
    coverage'(e) \sim B(\sum_s |\hat{C}(s)|, \frac{1}{|E|})
\end{equation}
So, the final output of TRACED is the set of edges for which the \textit{coverage} value is much higher than expected based on chance ($p<0.05$ in the binomial distribution).

% whether that step makes the algorithm heuristic. 
% whether we eventually produce a set of edges that explain ALL differences in the activation cascades of ALL sources — or whether there are some arbitrary thresholds at that step.
This final step makes the TRACED algorithm heuristic - the set of edges that we finally report is no longer guaranteed to explain all differences in the activation cascades of all sources. Nevertheless, the result captures edges that have influenced the activation cascades across many source nodes, and is therefore more reliable.

% In Section~\ref{sec:edge-metric-comparison}, we discuss some other potential metrics, and showed how edges with high \textit{coverage} also tend to perform well in other metrics. 

\section{A case study on major depressive disorder}
\label{sec:mdd-study}

The focus of this paper is on the analysis method presented in the previous section, rather than a specific application. 
To illustrate one potential application of this method, however, we summarize here the results of a comparison between a group of severe MDD patients and a group of healthy controls. 
The DTI data for this comparison was provided to us by Dr. Helen Mayberg's group and they were originally used in the PReDICT study~\cite{choi2014reconciling,dunlop2012predictors}. The PReDICT study was approved by Emory’s Institutional Review Board and the Grady Hospital Research Oversight Committee. 
We constructed structural brain networks applying probabilistic tractography on diffusion MRI scans of 90 MDD patients and 18 control subjects. The brain was parcellated into 396 regions (198 regions for each hemisphere) using the multi-modal cortical parcellation of Glasser et al.~\cite{glasser2016multi}, and the Brainnetome Atlas~\cite{fan2016human} for sub-cortical regions. We applied the linear threshold model and generated an activation cascade for each source node, and measured the cascade membership differences between the two groups. The threshold that we used ranges from 0.1 to 0.3 among different source nodes, and is determined for each source node as the one associated with most significant cascade membership differences. We then applied TRACED to identify the minimal set of connections that can explain the observed cascade differences.

Table~\ref{tab:root-cause-solutions} lists the connections that we identified as causal for the cascade membership differences between the two groups. These connections have a significant overlap with findings of earlier studies reporting MDD-related structural/functional changes. The connections identified as causal are adjacent to parts of Brodmann area 24~\cite{korgaonkar2014abnormal}, area 32~\cite{grieve2013widespread}, area 9~\cite{kerestes2012abnormal}, area 10~\cite{long2015disrupted}, and the orbitofrontal region~\cite{rajkowska1999morphometric}.  All of these regions have been reported to be pathologically relevant for MDD in earlier studies. Some of the reported connections are also in the default mode network (DMN), which has been shown to be heavily affected by  MDD \cite{korgaonkar2014abnormal}, with increased functional connectivity \cite{hamilton2015depressive}. We are going to further analyze this dataset and also compare our findings with those of other network analysis methods in a follow-up MDD-specific article.

\begin{table}[ht]
    \caption{The connections that can explain the cascade differences between a group of MDD patients and a group of Controls. The name of each node is based on the parcellation of Glasser et al.  \cite{glasser2016multi}, followed with a brief description of the location of that region (L: left hemisphere, R: right hemisphere). }
    \centering
    \begin{tabular}{llcrr}
    \toprule
    Node 1      & description               & --     & Node 2   & description             \\
    \midrule
    p24 (L)     & area-24 posterior         & --     & a24 (L)  & area-24 anterior        \\
    10v (L)     & area-10 ventral           & --     & 10pp (L) & medial polar area-10    \\
    a24 (L)     & area-24 anterior          & --     & 9m (L)   & area-9 medial           \\
    Pir (L)     & piriform olfactory cortex & --     & pOFC (L) & posterior OFC           \\
    13l (L)     & area-13 lateral           & --     & OFC (L)  & orbital frontal complex \\
    p32 (L)     & area-32 posterior         & --     & 10d (L)  & area-10 dorsal          \\
    p32 (L)     & area-32 posterior         & --     & 9m (L)   & area-9 medial           \\
    10v (R)     & area-10 ventral           & --     & 10pp (R) & polar 10p               \\
    pOFC (L)    & posterior OFC             & --     & 13l (L)  & area-13 larteral        \\
    10pp (L)    & medial polar area-10      & --     & OFC (L)  & orbital frontal complex \\
    p32 (L)     & area-32 posterior         & --     & 10pp (L) & medial polar area-10    \\
    \bottomrule
    \end{tabular}
    \label{tab:root-cause-solutions}
\end{table}

\section{Discussion}
\label{sec:discussion}

% Arguments that our method is a very good complement to the linear threshold models

% Compare our method with other methods
Various network analysis metrics and methods have been proposed in the past to compare structural brain networks. For instance, earlier work has investigated the differences between brain networks in terms of small-worldness~\cite{bassett2017small}, efficiency~\cite{berlot2016global}, and modularity~\cite{sporns2016modular}. At the node level, the clustering coefficient, participant coefficient, and different node centrality metrics (especially the betweenness centrality) have been widely adopted~\cite{qin2014abnormal,zhang2020rumination}. At the edge level, researchers have investigated the edges with significant weight differences and the subnetwork they form~\cite{korgaonkar2014abnormal}.

TRACED falls in the spectrum of the edge-level analysis, and the resulting set of connections is a subset of edges that have significant weight differences between the two groups. Additionally however TRACED also incorporates the information flow across the entire network in varied paths (because of all the source nodes considered). We aggregate this topological information across the entire network to describe the role that a specific network element (node or edge) plays in the network, and how that role is different between the two groups.

\begin{figure}[ht]
    \centering
    \includegraphics[width=\textwidth]{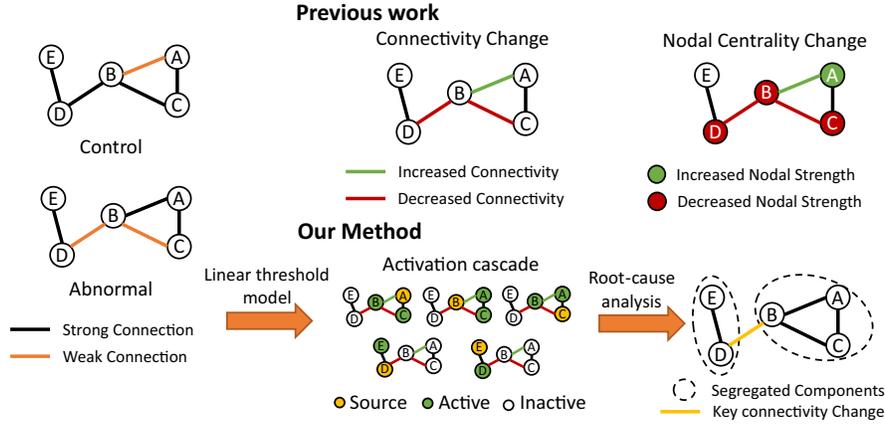}
    \caption{Earlier work has mostly focused on brain connectivity differences using graph-theoretic metrics (e.g. node centrality metrics). TRACED associates connectivity changes with their impact on information transfer in the brain. It measures the impact of such changes on activation cascade differences, and identifies the specific connections that cause these differences through root-cause analysis.}
    \label{fig:approach-comparison}
\end{figure}

%% Comparison with approaches based on connectivity
Fig.~\ref{fig:approach-comparison} illustrates typical node-level and edge-level network analysis metrics and compares them with TRACED. Compared to identifying solely edges with significant weight changes, TRACED associates a structural change (i.e., restoring the weight of a connection to its value in the other group) with functional changes (the node membership of the corresponding activation cascades). This is favorable for two reasons: it makes the results more interpretable, and less sensitive to variability across subjects. A significant difference in the weight of a connection between two networks may be simply due to subject variability. With TRACED, a connection is identified as causal not only based on its weight but also based on the topological role of that edge in the propagation of information (activation cascades) from different source nodes. 

%% Comparison with approaches based on network metrics
Compared to node-level analysis metrics, TRACED can provide higher spatial resolution because it identifies specific connections instead of entire brain regions. Additionally,  some network analysis metrics often make implicit assumptions about information transfer in the brain (e.g. the betweenness centrality metric assumes that information travels through shortest paths, while the communicability metric assumes that information follows random-walks). These assumptions may not be realistic (e.g. shortest path routing requires information about the complete network stored in every node). It is also harder to interpret these metrics in terms of their associated localities in the brain (e.g. a node may have much lower communicability in one group but what is the corresponding set of affected information pathways?). TRACED makes an explicit assumption about information transfer, namely activation cascades based on the linear threshold model, and it associates structural connectivity changes with corresponding functional changes, making the results more transparent and informative.

% \bibliographystyle{unsrt}
% \bibliographystyle{splncs04}
% \bibliography{references.bib} 

\begin{thebibliography}{10}
\providecommand{\url}[1]{\texttt{#1}}
\providecommand{\urlprefix}{URL }
\providecommand{\doi}[1]{https://doi.org/#1}

\bibitem{bassett2017small}
Bassett, D.S., Bullmore, E.T.: Small-world brain networks revisited. The
  Neuroscientist  \textbf{23}(5),  499--516 (2017)

\bibitem{berlot2016global}
Berlot, R., Metzler-Baddeley, C., Ikram, M.A., Jones, D.K., O’Sullivan, M.J.:
  Global efficiency of structural networks mediates cognitive control in mild
  cognitive impairment. Frontiers in aging neuroscience  \textbf{8}, ~292
  (2016)

\bibitem{choi2014reconciling}
Choi, K.S., Holtzheimer, P.E., Franco, A.R., Kelley, M.E., Dunlop, B.W., Hu,
  X.P., Mayberg, H.S.: Reconciling variable findings of white matter integrity
  in major depressive disorder. Neuropsychopharmacology  \textbf{39}(6),
  1332--1339 (2014)

\bibitem{dunlop2012predictors}
Dunlop, B.W., Binder, E.B., Cubells, J.F., Goodman, M.M., Kelley, M.E.,
  Kinkead, B., Kutner, M., Nemeroff, C.B., Newport, D.J., Owens, M.J., et~al.:
  Predictors of remission in depression to individual and combined treatments
  (predict): study protocol for a randomized controlled trial. Trials
  \textbf{13}(1),  1--18 (2012)

\bibitem{fan2016human}
Fan, L., Li, H., Zhuo, J., Zhang, Y., Wang, J., Chen, L., Yang, Z., Chu, C.,
  Xie, S., Laird, A.R., et~al.: The human brainnetome atlas: a new brain atlas
  based on connectional architecture. Cerebral cortex  \textbf{26}(8),
  3508--3526 (2016)

\bibitem{fischer2015altered}
Fischer, F.U., Wolf, D., Scheurich, A., Fellgiebel, A., Initiative, A.D.N.,
  et~al.: Altered whole-brain white matter networks in preclinical alzheimer's
  disease. NeuroImage: clinical  \textbf{8},  660--666 (2015)

\bibitem{fleischer2019graph}
Fleischer, V., Radetz, A., Ciolac, D., Muthuraman, M., Gonzalez-Escamilla, G.,
  Zipp, F., Groppa, S.: Graph theoretical framework of brain networks in
  multiple sclerosis: a review of concepts. Neuroscience  \textbf{403},  35--53
  (2019)

\bibitem{Fornito2012schizophrenia}
Fornito, A., Zalesky, A., Pantelis, C., Bullmore, E.T.: Schizophrenia,
  neuroimaging and connectomics. Neuroimage  \textbf{62}(4),  2296--2314 (2012)

\bibitem{glasser2016multi}
Glasser, M.F., Coalson, T.S., Robinson, E.C., Hacker, C.D., Harwell, J.,
  Yacoub, E., Ugurbil, K., Andersson, J., Beckmann, C.F., Jenkinson, M.,
  et~al.: A multi-modal parcellation of human cerebral cortex. Nature
  \textbf{536}(7615),  171--178 (2016)

\bibitem{grieve2013widespread}
Grieve, S.M., Korgaonkar, M.S., Koslow, S.H., Gordon, E., Williams, L.M.:
  Widespread reductions in gray matter volume in depression. NeuroImage:
  Clinical  \textbf{3},  332--339 (2013)

\bibitem{hamilton2015depressive}
Hamilton, J.P., Farmer, M., Fogelman, P., Gotlib, I.H.: Depressive rumination,
  the default-mode network, and the dark matter of clinical neuroscience.
  Biological psychiatry  \textbf{78}(4),  224--230 (2015)

\bibitem{hwang2013development}
Hwang, K., Hallquist, M.N., Luna, B.: The development of hub architecture in
  the human functional brain network. Cerebral Cortex  \textbf{23}(10),
  2380--2393 (2013)

\bibitem{kerestes2012abnormal}
Kerestes, R., Ladouceur, C.D., Meda, S., Nathan, P.J., Blumberg, H.P., Maloney,
  K., Ruf, B., Saricicek, A., Pearlson, G.D., Bhagwagar, Z., et~al.: Abnormal
  prefrontal activity subserving attentional control of emotion in remitted
  depressed patients during a working memory task with emotional distracters.
  Psychological Medicine  \textbf{42}(1),  29–40 (2012).
  \doi{10.1017/S0033291711001097}

\bibitem{korgaonkar2014abnormal}
Korgaonkar, M.S., Fornito, A., Williams, L.M., Grieve, S.M.: Abnormal
  structural networks characterize major depressive disorder: a connectome
  analysis. Biological psychiatry  \textbf{76}(7),  567--574 (2014)

\bibitem{llufriu2017structural}
Llufriu, S., Martinez-Heras, E., Solana, E., Sola-Valls, N., Sepulveda, M.,
  Blanco, Y., Martinez-Lapiscina, E.H., Andorra, M., Villoslada, P.,
  Prats-Galino, A., et~al.: Structural networks involved in attention and
  executive functions in multiple sclerosis. NeuroImage: Clinical  \textbf{13},
   288--296 (2017)

\bibitem{long2015disrupted}
Long, Z., Duan, X., Wang, Y., Liu, F., Zeng, L., Zhao, J.p., Chen, H.:
  Disrupted structural connectivity network in treatment-naive depression.
  Progress in Neuro-Psychopharmacology and Biological Psychiatry  \textbf{56},
  18--26 (2015)

\bibitem{qin2014abnormal}
Qin, J., Wei, M., Liu, H., Yan, R., Luo, G., Yao, Z., Lu, Q.: Abnormal brain
  anatomical topological organization of the cognitive-emotional and the
  frontoparietal circuitry in major depressive disorder. Magnetic resonance in
  medicine  \textbf{72}(5),  1397--1407 (2014)

\bibitem{rajkowska1999morphometric}
Rajkowska, G., Miguel-Hidalgo, J.J., Wei, J., Dilley, G., Pittman, S.D.,
  Meltzer, H.Y., Overholser, J.C., Roth, B.L., Stockmeier, C.A.: Morphometric
  evidence for neuronal and glial prefrontal cell pathology in major
  depression. Biological psychiatry  \textbf{45}(9),  1085--1098 (1999)

\bibitem{rehme2013cerebral}
Rehme, A.K., Grefkes, C.: Cerebral network disorders after stroke: evidence
  from imaging-based connectivity analyses of active and resting brain states
  in humans. The Journal of physiology  \textbf{591}(1),  17--31 (2013)

\bibitem{riva2018connectomic}
Riva-Posse, P., Choi, K., Holtzheimer, P.E., Crowell, A.L., Garlow, S.J.,
  Rajendra, J.K., McIntyre, C.C., Gross, R.E., Mayberg, H.S.: A connectomic
  approach for subcallosal cingulate deep brain stimulation surgery:
  prospective targeting in treatment-resistant depression. Molecular psychiatry
   \textbf{23}(4),  843--849 (2018)

\bibitem{shadi2020multisensory}
Shadi, K., Dyer, E., Dovrolis, C.: Multisensory integration in the mouse
  cortical connectome using a network diffusion model. Network Neuroscience
  \textbf{4}(4),  1030--1054 (2020)

\bibitem{sporns2016modular}
Sporns, O., Betzel, R.F.: Modular brain networks. Annual review of psychology
  \textbf{67},  613--640 (2016)

\bibitem{sporns2005connectome}
Sporns, O., Tononi, G., K{\"o}tter, R.: The human connectome: a structural
  description of the human brain. PLoS Comput Biol  \textbf{1}(4), ~e42 (2005)

\bibitem{stam2014modern}
Stam, C.J.: Modern network science of neurological disorders. Nature Reviews
  Neuroscience  \textbf{15}(10),  683--695 (2014)

\bibitem{stam2007small}
Stam, C.J., Jones, B., Nolte, G., Breakspear, M., Scheltens, P.: Small-world
  networks and functional connectivity in alzheimer's disease. Cerebral cortex
  \textbf{17}(1),  92--99 (2007)

\bibitem{van2015evidence}
Van~Hartevelt, T.J., Cabral, J., M{\o}ller, A., FitzGerald, J.J., Green, A.L.,
  Aziz, T.Z., Deco, G., Kringelbach, M.L.: Evidence from a rare case study for
  hebbian-like changes in structural connectivity induced by long-term deep
  brain stimulation. Frontiers in behavioral neuroscience  \textbf{9}, ~167
  (2015)

\bibitem{wen2017structural}
Wen, M.C., Heng, H.S., Hsu, J.L., Xu, Z., Liew, G.M., Au, W.L., Chan, L.L.,
  Tan, L.C., Tan, E.K.: Structural connectome alterations in prodromal and de
  novo parkinson's disease patients. Parkinsonism \& related disorders
  \textbf{45},  21--27 (2017)

\bibitem{zeng2012identifying}
Zeng, L.L., Shen, H., Liu, L., Wang, L., Li, B., Fang, P., Zhou, Z., Li, Y.,
  Hu, D.: Identifying major depression using whole-brain functional
  connectivity: a multivariate pattern analysis. Brain  \textbf{135}(5),
  1498--1507 (2012)

\bibitem{zhang2020rumination}
Zhang, R., Kranz, G.S., Zou, W., Deng, Y., Huang, X., Lin, K., Lee, T.M.:
  Rumination network dysfunction in major depression: A brain connectome study.
  Progress in Neuro-Psychopharmacology and Biological Psychiatry  \textbf{98},
  109819 (2020)

\end{thebibliography}

\clearpage
\appendix
\section{Appendix}

\subsection{Optimization of TRACED}
\label{sec:iterative-Optimization}
A key observation is that if adding a single edge $(x, y)$ into a solution set does not change the activation status of node $y$, we will inevitably need to  add additional edges pointing to $y$ to build a final solution. Otherwise, for a solution $C$ with $(x, y)$, we can always find a better solution $C' = C -\{(x, y)\} $ with $U'_C(s) = U'_{C'}(s)$.

Therefore, we can improve the original TRACED algorithm, by adding a collection of edges in each iteration, so that $U'_C(s)$ changes when we create a new partial solution. This way we can reduce the number of partial solutions that we create during the search for the optimal solution. How do we find the collection of edges that can cause the change in $U_C(s)$? We know that we focus on change of activation status of nodes in $U'_C(s) \triangle U(s)$, and so we can discuss the case of nodes $U(s) \setminus U'_C(s)$ and $U'_C(s) \setminus U(s)$ separately.
\begin{enumerate}
    \item For each node $v$ in $U(s) \setminus U'_C(s)$, we can check if there is an ensemble of edges from $U(s) \cap U'_C(s)$ pointing to this node, so that if we include the ensemble into the solution, $v$ would be active in the updated $U'_C(s)$. It is guaranteed that we can find at least one such collection of edges. Otherwise, we cannot explain why this $v$ could be active in $U(s)$. 
    \item For nodes in $U'_C(s) \setminus U(s)$, we can further find its subset $T_C(s)$ so that for each node $v \in T_C(s)$,  $\sum_{u \in U(s) \cap U'_C(s)} w(u, v) \geq \theta $. We can prove that $U'_C(s) \setminus U(s)$ will no longer be in $U'_C(s)$ if and only if we add an ensemble of edges for each node in $T_C(s)$ into $C$. If for a node $v$ in $T_C(s)$ we do not add edges connecting to $v$ into $C$, $v$ will remain active and present in $U'_C(s)$. If we add edges connecting to $v$ for every node $v$ in $T_C(s)$, none of the nodes in $U'_C(s) \setminus U(s)$ receive an activation more than $\theta$, so that they will no longer be active.
\end{enumerate}

With this modification, each partial solution $C$ corresponds to a state $U'_C(s)$, and it is guaranteed that there are no edges that can be removed from $C$ without changing that state. Therefore, all partial solutions corresponding to one state are equivalent, in terms of the edges that need to be added to the solution to reach another state. Therefore, we can construct a graph of solutions, where each node $x$ corresponds to a state, and each edge $(x, y, \{e_1, \dots\})$ corresponds to an ensemble of edges $\{e_1, \dots\}$ needed to be added into the partial solutions corresponding to state $x$ so that the new solution leads to state $y$. Such an edge is also weighted, with a weight that is equal to the number of edges in the collection. Notice that there can be multiple edges between two nodes, each corresponding to one collection of edges and may have a different weight different than other edges.

With such a graph of solutions, our goal is equivalent to finding the weighted shortest path between the initial state $U'(s)$ and the final state $U = U'_{\hat{C}}(s)$ in the graph. This is because the sum of the weights of edges along a path in the graph of solutions would be the number of actual edges we include in the final solution. We can find the shortest path using Dijkstra's algorithm since we have only positive weights. The major benefit of having this graph of solutions is that we can deal with the case of multiple optimal solutions more explicitly. They will be represented as multiple shortest paths from the initial state to the final state.

\end{document}